\documentclass[aps,prd,preprintnumbers,showpacs,superscriptaddress,nofootinbib,amsmath,amssymb,floats,floatfix,showkeys,notitlepage,longbibliography]{revtex4-1}
\usepackage[table]{xcolor} 
\usepackage{array} 
\definecolor{lightblue}{rgb}{0.62, 0.83, 0.96} 
\definecolor{lightgray}{rgb}{0.93, 0.97, 0.99} 
\usepackage{comment}
\usepackage{graphicx}
\usepackage{subfigure}
\usepackage{palatino}
\usepackage[commandnameprefix=always]{changes}
\usepackage{hyperref}
\hypersetup{colorlinks=true,linkcolor=red,urlcolor=red,citecolor=red}
\usepackage[toc,page]{appendix}
\usepackage[normalem]{ulem}
\usepackage{lipsum}
\usepackage{graphicx}
\usepackage{subfigure}
\usepackage{palatino}
\usepackage{float}
\usepackage{sans}
\usepackage{adjustbox}
\usepackage{latexsym}
\usepackage{amsmath}
\usepackage{amssymb}
\usepackage{amsfonts}
\usepackage{dcolumn}
\usepackage{bm}
\usepackage{tikz}
\usepackage{bigints}
\usepackage{array,tabularx,multirow,booktabs}
\usepackage[tracking=true]{microtype}
\SetTracking{}{500}
\SetTracking{encoding={*}, shape=sc}{40}
\allowdisplaybreaks
\usepackage{adjustbox}
\usepackage{latexsym}
\usepackage{amsmath}
\usepackage{amssymb}
\usepackage{amsfonts}
\usepackage{dcolumn}
\usepackage{bm}
\usepackage{tikz}
\usepackage{bigints}
\usepackage{array,tabularx,multirow,booktabs}
\usepackage[tracking=true]{microtype}
\usepackage{color}
\allowdisplaybreaks
\begin{document}

\title{Probing the Weak Gravity Conjecture: Novel Aschenbach Signatures in Superextremal Non-Linear Charged AdS Black Holes}

\author{Mohammad Reza Alipour}
\email{mohamad.alipour.1994@gmail.com,mr.alipour@stu.umz.ac.ir}
\affiliation{School of Physics, Damghan University, Damghan 3671645667, Iran}
\affiliation{Department of Physics, Faculty of Basic
Sciences, University of Mazandaran\\ P. O. Box 47416-95447, Babolsar, Iran}

\author{Mohammad Ali S Afshar}
\email{m.a.s.afshar@gmail.com}
\affiliation{Department of Physics, Faculty of Basic
Sciences, University of Mazandaran\\ P. O. Box 47416-95447, Babolsar, Iran}
\affiliation{School of Physics, Damghan University, Damghan 3671645667, Iran}
\affiliation{Canadian Quantum Research Center, 204-3002 32 Ave Vernon, BC V1T 2L7, Canada}

\author{Saeed Noori Gashti}
\email{saeed.noorigashti70@gmail.com,saeed.noorigashti@stu.umz.ac.ir}
\affiliation{School of Physics, Damghan University, Damghan 3671645667, Iran}

\author{Behnam Pourhassan}
\email{b.pourhassan@du.ac.ir}
\affiliation{School of Physics, Damghan University, Damghan 3671645667, Iran}
\affiliation{Center for Theoretical Physics, Khazar University, 41 Mehseti Street, Baku, AZ1096, Azerbaijan}

\author{Jafar Sadeghi}
\email{ pouriya@ipm.ir}
\affiliation{Department of Physics, Faculty of Basic
Sciences, University of Mazandaran\\ P. O. Box 47416-95447, Babolsar, Iran}
\affiliation{Canadian Quantum Research Center, 204-3002 32 Ave Vernon, BC V1T 2L7, Canada}
\begin{abstract}
This study investigates the nonlinear charged Anti-de Sitter (AdS) black hole solution within the framework of massive gravity, motivated by recent advancements linking the Weak Gravity Conjecture (WGC) to phenomena such as Weak Cosmic Censorship Conjecture (WCCC) and photon sphere dynamics. Building on these foundations, we focus on the Aschenbach effect—a relativistic phenomenon intricately tied to the geometry of photon spheres and known to occur in some special sub-extremal non rotating black holes. Our primary objective is to determine whether this effect persists not only up to the extremal limit but also beyond, into the superextremal regime, thus probing the stability and validity of black hole characteristics in these extreme conditions. By analyzing the nonlinear charged AdS black hole solutions in massive gravity, we demonstrate that the Aschenbach effect remains a robust feature across both extremal and superextremal configurations. This extension suggests that key relativistic signatures and the underlying spacetime structures associated with high-spin black holes continue to hold beyond classical boundaries. Our results provide new insights into the behavior of ultra-compact objects and highlight promising directions for exploring the limits of general relativity, as well as potential generalizations of the WGC and WCC in strong gravitational fields.
\end{abstract}

\date{\today}



\maketitle
\section{Introduction}\label{s1}
Quantum gravity remains one of the most intricate and captivating areas of modern physics, drawing interest from a wide range of scientific disciplines. Among the numerous approaches aimed at understanding this elusive framework, the swampland program has emerged as a significant line of inquiry. This research initiative is designed to identify the essential characteristics that any viable theory of quantum gravity must satisfy \cite{1,2,3,4,5,6,7,8}. A central tenet of the swampland perspective is that many low-energy effective field theories, despite appearing self-consistent, are incompatible with a fundamental theory of quantum gravity like string theory. The purpose of the swampland framework is to distinguish theories that can emerge from a consistent UV-complete theory (the so-called "landscape") from those that cannot (the "swampland") \cite{1,2,3,4,5,6,7,8}. The motivation for this program stems from insights across several foundational areas, including the thermodynamics of black holes, holography via the AdS/CFT correspondence, and detailed constructions within string theory. Through these diverse threads, the swampland initiative seeks to shed light on the deep principles underlying quantum gravity, and their implications for both cosmological models and particle physics \cite{a,b,f,g,h,i,j,k,l,m,n,o,p,q,r,s,t,u,v,w,x,y,z,aa,bb,cc,dd,ee,ff,gg,hh,ii,jj,kk,ll,mm,nn,oo,pp,qq,rr,ss,tt,uu,vv,ww,xx,yy,zz,aaa,bbb,ccc,ddd,eee,fff,ggg,hhh,iii,jjj,kkk,lll,rrr,sss,ttt,uuu,vvv,www,xxx,zzz,zzzz}.\\

A foundational principle arising from the swampland conjectures is the incompatibility of global symmetries with a theory of quantum gravity. In contrast, gauge symmetries are allowed. This distinction has profound consequences, one of which is embodied in the WGC. The WGC asserts that in any consistent quantum gravitational theory, there must exist particles whose charge-to-mass ratio exceeds unity, ensuring that gravity is not the strongest force. The WGC, among other related conjectures, provides criteria to judge whether a given low-energy theory can arise from a UV-complete quantum gravitational model. Those interested in a deeper understanding of the swampland framework are encouraged to consult the broad and growing literature, which covers various conjectures and their applications. The swampland approach has also found relevance in topics such as black brane solutions, inflationary cosmology, and generalized thermodynamics, making it a fertile ground for ongoing investigation. Much of the conceptual groundwork for this field has been laid by earlier advances in string theory, black hole physics, and the study of holographic dualities. These historical developments have furnished researchers with the theoretical tools necessary to rigorously explore quantum gravity \cite{9,10}.\\

Another significant idea in gravitational physics is the Weak Cosmic Censorship Conjecture (WCCC), originally proposed by Roger Penrose. This conjecture addresses the nature of singularities in general relativity, suggesting that they are always hidden within event horizons, thereby maintaining the deterministic structure of spacetime. However, difficulties arise when the WCCC is considered in tandem with the WGC, particularly in scenarios involving Reissner–Nordström (RN) black holes, which are charged solutions to the Einstein–Maxwell equations. If the charge of such a black hole exceeds its mass, the WCCC is violated. Conversely, if the black hole decays to an extremal state where its charge equals its mass, the resulting decay products imply a charge-to-mass ratio greater than one, satisfying the WGC but challenging the WCCC. This apparent tension raises fundamental questions about their compatibility. Recent theoretical investigations have shown that certain factors—such as the presence of dark matter and a nonzero cosmological constant—can mitigate these issues. In particular, these elements can prevent the overcharging of black holes under specific conditions that balance their respective influences. This opens new pathways to examine potential resolutions to the conflict between the WGC and the WCCC \cite{10000,11,12}.\\

The Aschenbach effect is a striking relativistic phenomenon that arises in the environment surrounding rapidly rotating black holes. This effect, first identified through computational models by Bernd Aschenbach in 2004, reveals an unexpected drop in the orbital velocity of particles moving in close proximity to a Kerr black hole's event horizon. Unlike traditional orbital behavior predicted by Newtonian or Keplerian mechanics, this unusual velocity dip occurs just before a particle reaches the point of no return, challenging long-held assumptions about motion in strong gravitational fields. What sets the Aschenbach effect apart is that it does not rely on electromagnetic processes or fluid dynamics. Instead, it emerges purely from the mathematical structure of the Kerr metric, which describes the spacetime geometry around a rotating black hole. This makes the phenomenon fundamentally different from quasi-periodic oscillations (QPOs), which typically require more complex modeling involving magnetohydrodynamic effects. Several key elements underlie the appearance of this effect. One is the innermost stable circular orbit (ISCO), which marks the inner boundary where a particle can maintain a stable orbit around a black hole \cite{8500,8501,8502,8503,8504,8505,8505,8506,8507,8508,8509,8510,8511,8512,8513,8514,8515,8516,8517,8518,8519,8520,8521,8522}. Inside this radius, any small disturbance causes the object to spiral inward. The ISCO’s location is highly sensitive to the black hole’s angular momentum and differs for particles moving in the same direction as the black hole's spin versus those moving against it. Another essential factor is frame dragging, a unique feature of rotating black holes where the spin of the central mass pulls the surrounding spacetime along with it. This rotation-altered spacetime significantly affects the paths and speeds of nearby particles, especially in regions closest to the event horizon. Additionally, strong gravitational fields near the event horizon drastically reshape the spacetime fabric. In these extreme conditions, the interaction between gravity and rotation encoded in the Kerr solution becomes the dominant influence on particle dynamics, leading to the velocity reversal that defines the Aschenbach effect.The Aschenbach effect represents a purely relativistic and kinematic phenomenon linked to the strong-field behavior of rotating black holes. Its study not only deepens our understanding of black hole environments but also provides a powerful tool for testing Einstein’s theory of gravity where it is pushed to its limits \cite{8500,8501,8502,8503,8504,8505,8505,8506,8507,8508,8509,8510,8511,8512,8513,8514,8515,8516,8517,8518,8519,8520,8521,8522}.\\
One particularly intriguing insight emerging from recent investigations is the possibility of replicating the non-monotonic behavior observed in the angular velocity profile of a rotating Kerr black hole through non-rotating black hole configurations. These studies demonstrate that, even in the absence of frame-dragging effects typically associated with rotation, angular velocity profiles can still exhibit non-uniform characteristics. This phenomenon arises solely due to the peculiar curvature of spacetime—provided that the geometry permits the formation of suitable potential extrema.  The term “potential extrema” here refers to the existence of local or global minima in the effective potential constructed from the spacetime metric. From a geodesic perspective, such a minimum corresponds to the emergence of a stable photon sphere located outside the event horizon \cite{46m,403,405,406}.
The primary motivation for this study arises from prior investigations that have provided evidence supporting the WGC through frameworks such as WCCC and the analysis of photon spheres. Building on this foundation, our objective is to examine whether the Aschenbach effect—a phenomenon closely associated with the structure and dynamics of photon spheres—continues to manifest consistently not only up to the extremal limit of rotating black holes, but even beyond, into the superextremal regime. The Aschenbach effect has been extensively studied in the context of sub-extremal black holes, where it is recognized as a distinctive relativistic signature linked to the geometry of spacetime \cite{46m,403,405,406}. Given that photon spheres play a critical role in shaping the conditions under which this effect emerges, our work aims to explore whether the underlying physical features responsible for the effect persist as the black hole approaches, and even slightly exceeds, extremality. In this paper, we demonstrate that the Aschenbach effect—previously observed in sub-extremal black holes within the studied model \cite{403}—remains valid and theoretically robust even as the system approaches or enters the regime of over-extremal configurations. This suggests that the characteristic properties of black holes, particularly those associated with high-spin configurations, may extend into regimes that challenge the classical boundaries of black hole solutions. Our findings indicate that the persistence of the Aschenbach effect beyond the extremal limit opens up intriguing possibilities for further research. In particular, it invites a deeper examination of how relativistic phenomena behave in transitional or borderline cases, and whether such behaviors could offer new insights into the nature of ultra-compact objects, the limits of general relativity, and potential extensions of the WGC or WCC in extreme gravitational environments.\\

The structure of this paper is organized as follows: In Section II, we introduce the theoretical framework that describes nonlinear charged Anti-de Sitter (AdS) black holes within the context of massive gravity. This section outlines the key modifications to the classical black hole solutions and discusses the parameters that characterize the model. Section III focuses on the analysis of two significant topics: the properties of photon spheres (PSs) and the Weak Gravity Conjecture (WGC), examining their mutual consistency and physical implications. In Section IV, we investigate the Aschenbach-like effect and its relation to the extremality condition of black holes, exploring how these phenomena manifest within the modified gravitational framework. Finally, Section V presents a summary of the main results and conclusions, along with a discussion of their broader significance and potential directions.

\section{The Model}\label{sec2}
In this section, we review the a nonlinear charged AdS black hole in massive gravity, a solution recently derived in \cite{9000,9001}. This black hole arises by solving the equations of motion of a four-dimensional system where massive gravity is coupled to a nonlinear electromagnetic field in an AdS spacetime background. The action describing this system is expressed as:
\begin{equation}\label{m1}
S = \int d^4x \sqrt{-g} \left[ \frac{1}{16\pi} \left( R - 2\Lambda + m^2 \sum_{i=1}^4 c_i U_i(g,f) \right) - \frac{1}{4\pi} L(F) \right],
\end{equation}
where $R$ denotes the scalar curvature of spacetime, and $\Lambda$ is the negative cosmological constant related to the AdS radius $l$ by $\Lambda = -\frac{3}{l^2}$. The parameter $m$ represents the graviton mass, while the $c_i$ are coupling constants associated with massive gravity. The fixed reference metric $f$ enters the action through symmetric polynomials $\mathcal{U}_i$, which are constructed from the eigenvalues of the matrix
\begin{equation}\label{m2}
\mathcal{K}^\mu_\nu = \sqrt{g^{\mu\lambda} f_{\lambda\nu}},
\end{equation}
with the polynomials defined as:
\begin{equation}\label{m3}
\begin{aligned}
\mathcal{U}_1 &= [\mathcal{K}], \\
\mathcal{U}_2 &= [\mathcal{K}]^2 - [\mathcal{K}^2], \\
\mathcal{U}_3 &= [\mathcal{K}]^3 - 3[\mathcal{K}][\mathcal{K}^2] + 2[\mathcal{K}^3], \\
\mathcal{U}_4 &= [\mathcal{K}]^4 - 6[\mathcal{K}]^2 [\mathcal{K}^2] + 8[\mathcal{K}][\mathcal{K}^3] + 3[\mathcal{K}^2]^2 - 6[\mathcal{K}^4],
\end{aligned}
\end{equation}
where square brackets denote the trace operator, e.g., $[\mathcal{K}] = \mathcal{K}^\mu_\mu$. The nonlinear electromagnetic Lagrangian $L(F)$ depends on the field invariant
\begin{equation}\label{m4}
F = \frac{1}{4} F_{\mu\nu} F^{\mu\nu},
\end{equation}
with the electromagnetic field strength tensor defined by $F_{\mu\nu} = \partial_\mu A_\nu - \partial_\nu A_\mu$. It takes the form
\begin{equation}\label{m5}
L(F) = F e^{-\frac{k}{2Q} (2Q^2 F)^{\frac{1}{4}}},
\end{equation}
where $Q$ is the total charge of the system, $k$ is a fixed parameter relating the charge and the mass via $Q^2 = M k$. Assuming a spherically symmetric and static spacetime, the fixed reference metric is chosen as \cite{9002}
\begin{equation}\label{m6}
f_{\mu\nu} = \mathrm{diag}(0, 0, c^2, c^2 \sin^2 \theta),
\end{equation}
where $c$ is a positive constant. From the variation of the action, the equations of motion are derived as:
\begin{equation}\label{m7}
G^\nu_\mu - \left( \frac{3}{l^2} + m^2 \left( \frac{c c_1}{r} + \frac{c^2 c_2}{r^2} \right) \right) \delta^\nu_\mu = 2 \left( \frac{\partial L(F)}{\partial F} F_{\mu\rho} F^{\nu\rho} - \delta^\nu_\mu L(F) \right),
\end{equation}
alongside the generalized Maxwell equations:
\begin{equation}\label{m8}
\nabla_\mu \left( \frac{\partial L(F)}{\partial F} F^{\nu\mu} \right) = 0, \quad \nabla_\mu *F^{\nu\mu} = 0.
\end{equation}
Solving these equations yields a static, spherically symmetric black hole solution characterized by mass $M$ and magnetic charge $Q$. The metric is:
\begin{equation}\label{m9}
ds^2 = -f(r) dt^2 + \frac{dr^2}{f(r)} + r^2 d\Omega_2^2,
\end{equation}
with the electromagnetic field strength tensor components
\begin{equation}\label{m10}
F_{\mu\nu} =(\delta^\theta_\mu\delta^\phi_\nu-\delta^\theta_\nu\delta^\phi_\mu) B(r,\theta) .
\end{equation}
where $B(r,\theta) = Q \sin \theta$. The metric function $f(r)$ reads:
\begin{equation}\label{m11}
f(r) = 1 - \frac{2M}{r} e^{-\frac{k}{2r}} + \frac{r^2}{l^2} + m^2 \left( \frac{c c_1 r}{2} + c^2 c_2 \right).
\end{equation}
The term $1 + m^2 c^2 c_2$ acts as an effective horizon curvature, which can represent spherical ($>0$), flat (=0), or hyperbolic ($<0$) geometries depending on the sign and magnitude of $c_2$. When $M = Q = 0$, the vacuum solution reduces to
\begin{equation}\label{m12}
f(r) = 1 + \frac{r^2}{l^2} + m^2 \left( \frac{c c_1 r}{2} + c^2 c_2 \right).
\end{equation}
At large distances where $k/r \ll 1$, the metric function approximates
\begin{equation}\label{m13}
f(r) \approx 1 - \frac{2M}{r} + \frac{Q^2}{r^2} + \frac{r^2}{l^2} + m^2 \left( \frac{c c_1 r }{2}+ c^2 c_2 \right),
\end{equation}
indicating that asymptotically, the black hole behaves like a four-dimensional Reissner–Nordström AdS black hole modified by massive gravity terms. Finally, the black hole mass $M$ can be expressed in terms of the event horizon radius $r_h$ and the pressure $P = -\frac{\Lambda}{8\pi} = \frac{3}{8\pi l^2}$ as
\begin{equation}\label{m14}
M = \frac{r_h}{2} e^{\frac{k}{2r_h}} \left( 1 + \frac{8\pi P r_h^2}{3} + m^2 \left( \frac{c c_1 r_h}{2} + c^2 c_2 \right) \right).
\end{equation}
The first law of black hole thermodynamics then relates changes in mass to variations in entropy $S$, pressure $P$, and massive gravity parameters $c_i$. In the context of massive gravity, the coupling constants $c_1$ and $c_2$ are generally variable and treated as thermodynamic parameters \cite{9003,9004,9005}. Their conjugate quantities are denoted by $c_1$ and $c_2$, respectively. The temperature of the black hole is determined from the surface gravity evaluated at the event horizon radius $r_h$, and can be written as
\begin{equation}\label{m15}
T = \frac{f'(r_h)}{4\pi} = \left(2r_h - \frac{k}{3}\right) P + \frac{2r_h - k}{8\pi r_h^2} \left[ 1 + m^2 \left( \frac{c c_1 r_h}{2} + c^2 c_2 \right) \right] + \frac{m^2 c c_1}{8\pi}.
\end{equation}
The entropy $S$, thermodynamic volume $V$, and the conjugate quantities $C_1$, $C_2$ follow from the first law of thermodynamics. The entropy can be found by integrating
\begin{equation}\label{m16}
S = \int \frac{1}{T} \left(\frac{\partial M}{\partial r_h}\right) dr_h,
\end{equation}
which yields
\begin{equation}\label{m17}
S = \pi r_h^2 \left(1 + \frac{k}{2r_h}\right) e^{\frac{k}{2r_h}} - \frac{\pi k^2}{4} \operatorname{Ei}\left(\frac{k}{2r_h}\right),
\end{equation}
where $\operatorname{Ei}(x)$ is the exponential integral function defined by
\begin{equation}\label{m18}
\operatorname{Ei}(x) = -\int_{-x}^\infty \frac{e^{-t}}{t} dt.
\end{equation}
The thermodynamic volume associated with the pressure $P$ is
\begin{equation}\label{m19}
V = \left(\frac{\partial M}{\partial P}\right)_{S,c_1,c_2} = \frac{4\pi r_h^3}{3} e^{\frac{k}{2r_h}},
\end{equation}
while the conjugates to the couplings $c_1$ and $c_2$ are
\begin{equation}\label{m20}
C_1 = \left(\frac{\partial M}{\partial c_1}\right)_{S,P,c_2} = \frac{m^2 c r_h^2}{4} e^{\frac{k}{2r_h}}, \quad
C_2 = \left(\frac{\partial M}{\partial c_2}\right)_{S,P,c_1} = \frac{m^2 c^2 r_h}{2} e^{\frac{k}{2r_h}}.
\end{equation}
Notably, the black hole entropy deviates from the usual area law only due to the nonlinear electrodynamics effects. For large horizon radius $r_h$ satisfying $k / r_h \ll 1$, the entropy expansion approximates
\begin{equation}\label{m21}
S = \pi k^{2} \left( \left(\frac{r_{+}}{k}\right)^{2} + \frac{r_{+}}{k} + \frac{3 - 2\gamma}{8} - \frac{1}{4} \ln \left(\frac{k}{2r_{+}}\right) + O\left(\frac{k}{r_{+}}\right) \right),
\end{equation}
where $\gamma \approx 0.5772$ is Euler–Mascheroni constant. This confirms the approximate validity of the area law for large black holes. The heat capacity at constant pressure $C_P$ is computed by
\begin{equation}\label{m22}
C_P = T \left( \frac{\partial S}{\partial T} \right)_P = \frac{\partial M / \partial r_h}{\partial T / \partial r_h} = \pi r_h^2 \frac{ h_1(r_h)}{6h_2(r_h)} e^{\frac{k}{2r_h}},
\end{equation}
with
\begin{equation}\label{m23}
\begin{aligned}
h_1(r_h) &= 16 \pi P r_h^2 (6 r_h - k) + 3 m^2 c c_1 r_h (4 r_h - k) + 6 (1 + m^2 c^2 c_2)(2 r_h - k), \\
h_2(r_h) &= 8 \pi P r_h^3 - \left( 1 + m^2 c^2 c_2 - \frac{m^2 c c_1 k}{4} \right) r_h + k (1 + m^2 c^2 c_2).
\end{aligned}
\end{equation}
From the temperature expression, the equation of state linking pressure $P$, temperature $T$, and horizon radius $r_h$ is
\begin{equation}\label{m24}
P = \frac{T}{2 r_h - \frac{k}{3}} + \frac{2(1 + m^2 c^2 c_2)(k - 2 r_h) + m^2 c c_1 r_h (k - 4 r_h)}{16 \pi r_h^2 (2 r_h - \frac{k}{3})}.
\end{equation}
Here, $r_h$ is regarded as a function of the thermodynamic volume $V$, defined earlier. When the effective horizon curvature $1 + m^2 c^2 c_2$ is positive, the system exhibits a critical point corresponding to an inflection point in the $P - r_h$ isotherm:
\begin{equation}\label{m25}
\left( \frac{\partial P}{\partial r_h} \right)_T = 0, \quad \left( \frac{\partial^2 P}{\partial r_h^2} \right)_T = 0.
\end{equation}
Solving these conditions gives the critical radius, temperature, and pressure as
\begin{equation}\label{m26}
r_c = \frac{6 k (1 + m^2 c^2 c_2)}{4 (1 + m^2 c^2 c_2) - m^2 c c_1 k},
\end{equation}
\begin{equation}\label{m27}
T_c = \frac{13 (1 + m^{2} c_{2} c^{2})}{81 \pi k} 
+ \frac{37 m^{2} c c_{1}}{216 \pi} 
+ \frac{(m^{2} c c_{1})^{2} k\left(1 + m^{2} c_{2} c^{2} + \frac{m^{2} c c_{1} k}{96}\right)}{ 108 \pi (1 + m^{2} c_{2} c^{2})^{2}},
\end{equation}
\begin{equation}\label{m28}
P_c = \frac{(1 + m^2 c^2 c_2 - \frac{m^2 c c_1 k}{4})^3}{54 \pi k^2 (1 + m^2 c^2 c_2)^2}.
\end{equation}
Below this critical temperature $T_c$, the black hole undergoes a first-order phase transition between small and large black hole states, analogous to the liquid-gas transition in a van der Waals fluid. This phase transition ceases to exist when $T > T_c$.
 \begin{figure}[h!]
 \begin{center}
 \subfigure[]{
 \includegraphics[height=4cm,width=5cm]{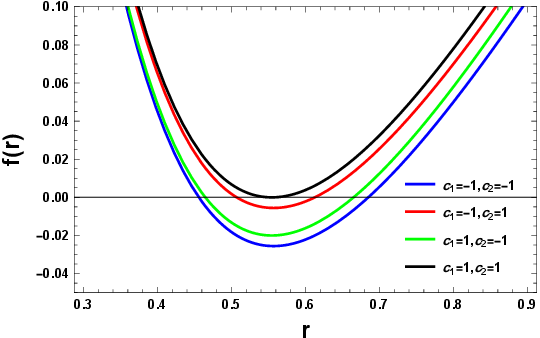}
 \label{fig1a}}
  \subfigure[]{
 \includegraphics[height=4cm,width=5cm]{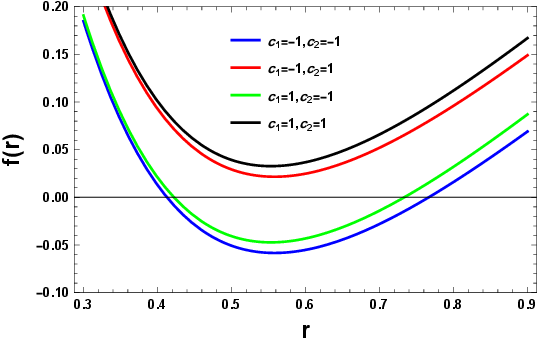}
 \label{fig1b}}
 \subfigure[]{
 \includegraphics[height=4cm,width=5cm]{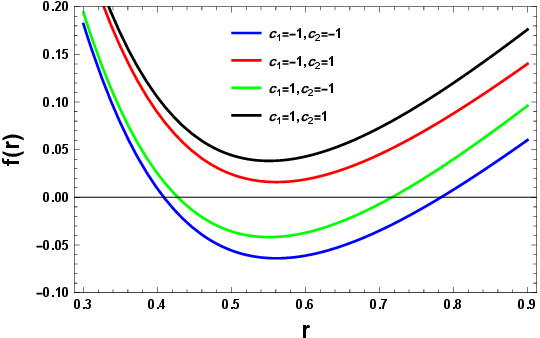}
 \label{fig1b}}
 \caption{\small{ The metric function with respect to $\ell=3$, $Q=1$, and $M=0.84$.
 (a) $m=0.1$, and $c=1$. (b) $m=0.1$, and $c=2$. (c) $m=0.2$, and $c=1$ . }}
 \label{fig1}
 \end{center}
 \end{figure}
We rewrite the mass and temperature so that we have 
\begin{equation}\label{eq2}
\begin{split}
M=\bigg[Q^2\bigg]\times\bigg[2 r_h W\left(\frac{2 Q^2 \ell ^2}{ c c_1  m^2 r_h^3 \ell ^2+2r_h^2c^2 c_2  m^2 \ell ^2+2 r_h^4+2r_h^2 \ell ^2}\right)\bigg]^{-1}
\end{split}
\end{equation}
and
\begin{equation}\label{eq2}
\begin{split}
&T=\frac{1}{2} c_1 c m^2+\frac{2 r_h}{\ell ^2}+\frac{2 \ell ^2 \left(c^2 c_2 m^2+1\right)+c c_1 m^2 \ell ^2 r_h+2 r_h^2}{2 \ell ^2 r_h}\\&\times\bigg(1-W\bigg[\frac{2 Q^2 \ell ^2}{2 c^2 c_2 m^2 \ell ^2 r_h^2+c c_1 m^2 \ell ^2 r_h^3+2 \ell ^2 r_h^2+2 r_h^4}\bigg] \bigg)
\end{split}
\end{equation}

\begin{table}[h!]
\centering
\setlength{\arrayrulewidth}{0.4mm} 
\setlength{\tabcolsep}{2.8pt} 
\arrayrulecolor[HTML]{000000} 
\begin{tabular}{|>{\centering\arraybackslash}m{2cm}|>{\centering\arraybackslash}m{2cm}|>{\centering\arraybackslash}m{2cm}|>{\centering\arraybackslash}m{2cm}|>{\centering\arraybackslash}m{2cm}|>{\centering\arraybackslash}m{2cm}|>{\centering\arraybackslash}m{2cm}|>{\centering\arraybackslash}m{2cm}|}
\hline
\rowcolor[HTML]{9FC5E8} 
\textbf{$c_1$} & \textbf{$c_2$} & \textbf{$Q_e$} & \textbf{$r_{e}$} & \textbf{$M_{e}$} & \textbf{$(Q/M)_{e}>1$} & \textbf{$WGC-WCCC$}\\ \hline
\rowcolor[HTML]{EAF4FC} 
-1 & -1 & 1 & 0.5622 & 0.8344 & 1.1985 & $\checkmark$ \\ \hline
-1 & 1 & 1 & 0.5580 & 0.8422 & 1.1873 & $\checkmark$ \\ \hline
\rowcolor[HTML]{EAF4FC}
1 & -1 & 1 & 0.5581 & 0.8366 & 1.1953 & $\checkmark$ \\ \hline
1 & 1 & 1 & 0.5545 & 0.8444 & 1.1842 & $\checkmark$ \\ \hline

\end{tabular}
\caption{The condition for consistency WGC-WCCC for $\ell=3$, $m=0.1$ and $c=1$.}
\label{P1}
\end{table}
\subsection{PSs and WGC}
\begin{figure}[h!]
 \begin{center}
 \subfigure[]{
 \includegraphics[height=4cm,width=5cm]{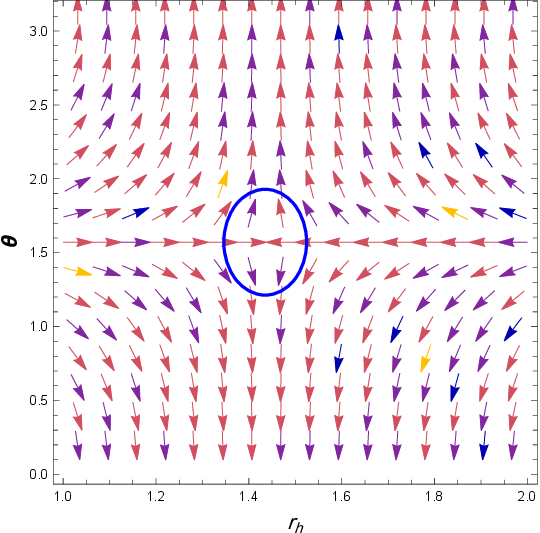}
 \includegraphics[height=4cm,width=5cm]{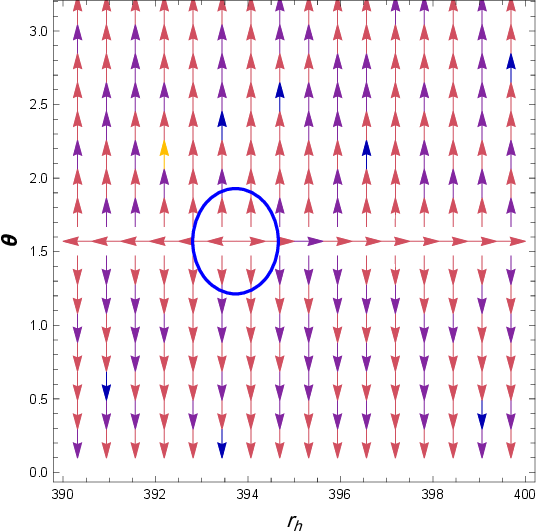}
 \label{fig2a}}
  \subfigure[]{
 \includegraphics[height=4cm,width=5cm]{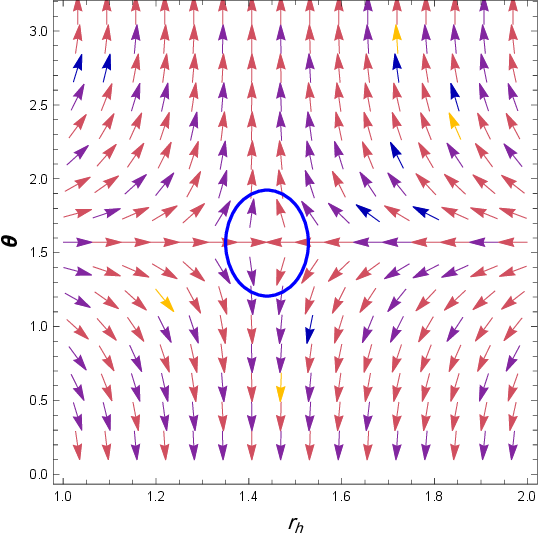}
 \includegraphics[height=4cm,width=5cm]{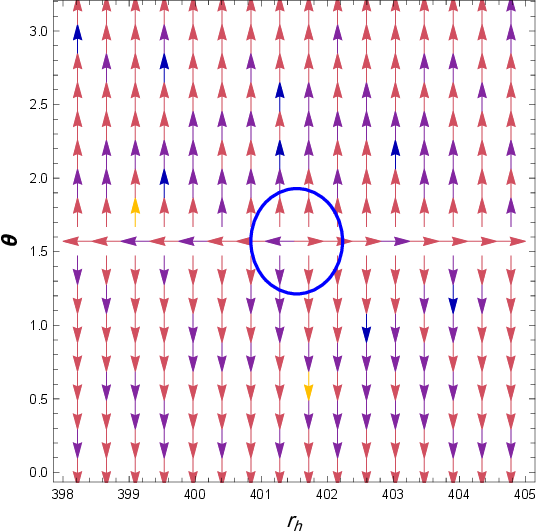}
 \label{fig2b}}
 \caption{\small{ The normal vector in the \( (r, \theta) \) plane associated with the photon spheres with respect to $\ell=3$, $m=0.1$, $c=1$, and $c_1=-1$.
 (a) $Q_e=1$, $M_e=0.8344$ and $c_2=-1$. (b) $Q_e=1$, $M_e=0.8422$ and $c_2=1$.  }}
 \label{fig2}
 \end{center}
 \end{figure}

\begin{figure}[h!]
 \begin{center}
 \subfigure[]{
 \includegraphics[height=4cm,width=5cm]{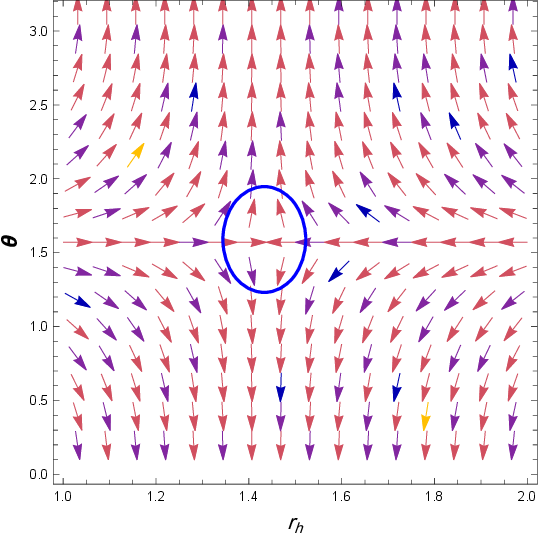}
 \label{fig3a}}
  \subfigure[]{
 \includegraphics[height=4cm,width=5cm]{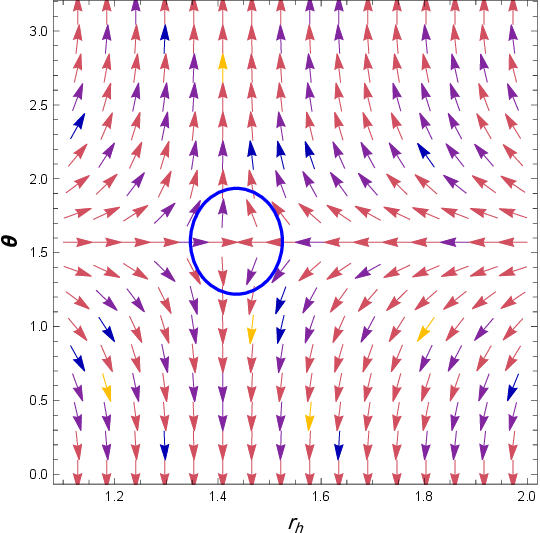}
 \label{fig3b}}
 \caption{\small{ The normal vector in the \( (r, \theta) \) plane associated with the photon spheres with respect to $\ell=3$, $m=0.1$, $c=1$, and $c_1=1$.
 (a) $Q_e=1$, $M_e=0.8366$ and $c_2=-1$. (b) $Q_e=1$, $M_e=0.8444$ and $c_2=1$.  }}
 \label{fig3}
 \end{center}
 \end{figure}

\begin{figure}[h!]
 \begin{center}
 \subfigure[]{
 \includegraphics[height=4cm,width=5cm]{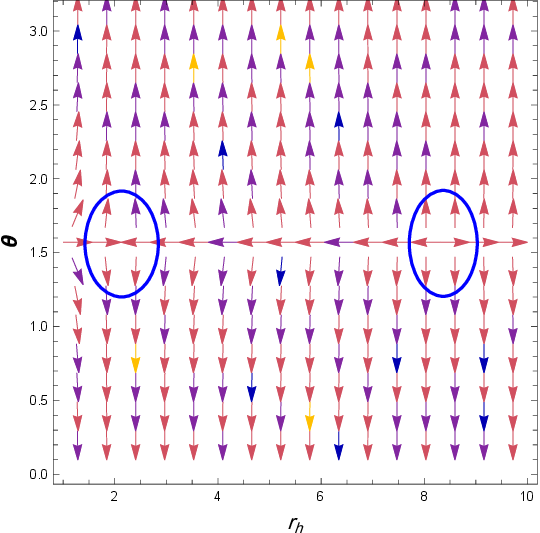}
 \label{fig4a}}
  \subfigure[]{
 \includegraphics[height=4cm,width=5cm]{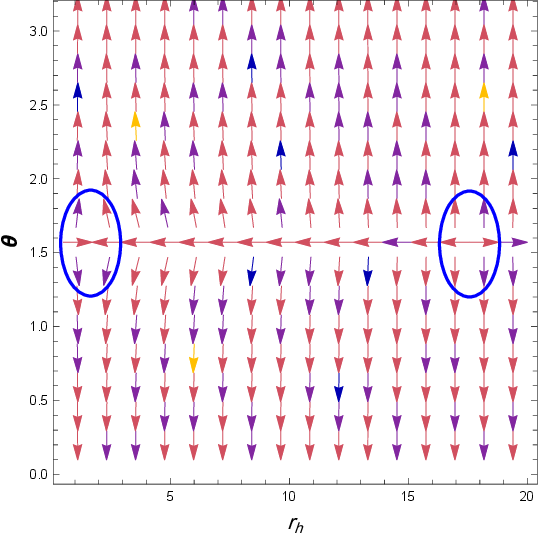}
 \label{fig4b}}
 \caption{\small{ The normal vector in the \( (r, \theta) \) plane associated with the photon spheres with respect to $\ell=3$, $m=0.5$, $c=1$, and $c_1=-1$.
 (a) $Q_e=1$, $M_e=0.6981$ and $c_2=-1$. (b) $Q_e=1$, $M_e=0.9086$ and $c_2=1$.  }}
 \label{fig4}
 \end{center}
 \end{figure}

\begin{figure}[h!]
 \begin{center}
 \subfigure[]{
 \includegraphics[height=4cm,width=5cm]{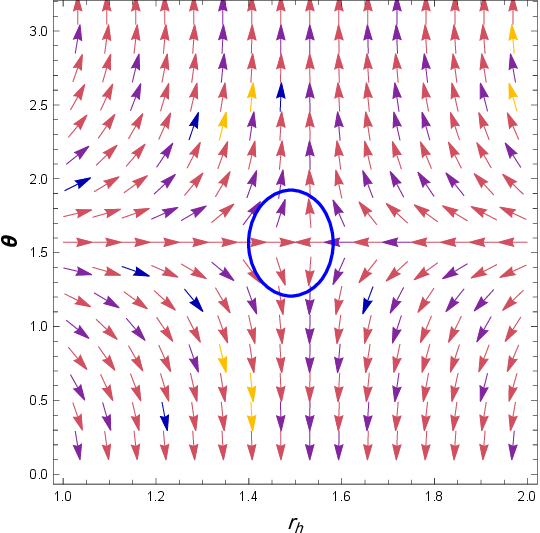}
 \label{fig5a}}
  \subfigure[]{
 \includegraphics[height=4cm,width=5cm]{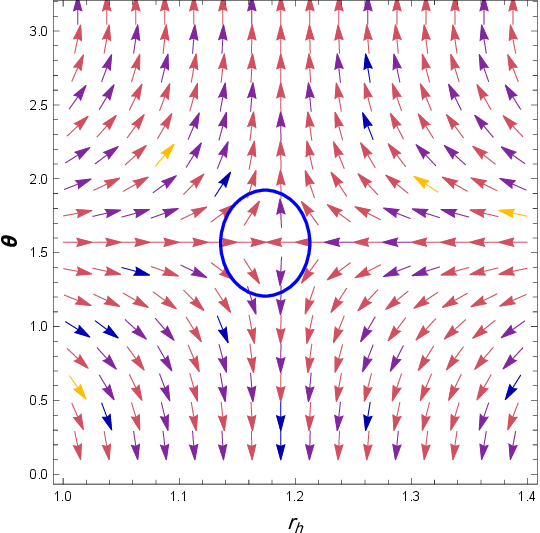}
 \label{fig5b}}
 \caption{\small{ The normal vector in the \( (r, \theta) \) plane associated with the photon spheres with respect to $\ell=3$, $m=0.5$, $c=1$, and $c_1=1$.
 (a) $Q_e=1$, $M_e=0.7668$ and $c_2=-1$. (b) $Q_e=1$, $M_e=0.9546$ and $c_2=1$.  }}
 \label{fig5}
 \end{center}
 \end{figure}

To investigate the characteristics of photon spheres in black hole geometries, we begin by examining the previously introduced metric (refer to Eq.\eqref{m11}), drawing on earlier research \cite{45'm,45mmm,45m,46m,47m,48m,49m,50m}. Our goal is to study the trajectories of lightlike particles and determine the radius at which photons can maintain circular orbits, known as photon spheres, using an effective potential framework. Due to the spacetime’s symmetry under reflection across the equatorial plane (a $Z_2$ symmetry), analysis is restricted to the plane defined by $\theta = \pi/2$, which reduces complexity without loss of generality. Within this plane, the radial motion of photons is governed by:
\begin{equation}\label{ph2}
\dot{r}^2 + V_{\text{eff}} = 0,
\end{equation}
where the effective potential $V_{\text{eff}}$ takes the form:
\begin{equation}\label{ph3}
V_{\text{eff}} = g(r) \left(\frac{L^2}{r^2} - \frac{E_p^2}{f(r)}\right).
\end{equation}
Here, $E_p$ and $L$ represent the constants of motion related to the photon's energy and angular momentum, which arise from the spacetime’s Killing symmetries in time and angular coordinates. For a stable circular photon orbit at radius $r_{\text{ps}}$, the effective potential must meet the criteria:
\begin{equation}\label{ph4}
V_{\text{eff}} = 0, \quad \text{and} \quad \frac{dV_{\text{eff}}}{dr} = 0.
\end{equation}
Solving these simultaneously yields the condition:
\begin{equation}\label{ph5}
\left(\frac{f(r)}{r^2}\right)' \bigg|_{r=r_{\text{ps}}} = 0,
\end{equation}
where the prime denotes differentiation with respect to the radial coordinate $r$. The stability of these circular paths is determined by the second derivative of $V_{\text{eff}}$: a negative value signals instability, while a positive value indicates stable orbits. Explicitly differentiating provides:
\begin{equation}\label{ph6}
2 r f(r) - r^2 f'(r) = 0.
\end{equation}
At the black hole event horizon $r_h$, the metric function satisfies $f(r_h) = 0$, which nullifies the first term. Since typically $f'(r_h) \neq 0$, this implies that photon spheres must reside outside the horizon. Nonetheless, in the extremal scenario, both $f(r_h)$ and its derivative vanish simultaneously, allowing the photon sphere radius to coincide with the degenerate horizon. Beyond this classical geodesic analysis, a topological interpretation is adopted to describe photon spheres. Each photon sphere configuration is assigned a topological charge that reflects its dynamical properties and stability. To this end, we define a scalar function:
\begin{equation}\label{ph7}
H(r, \theta) =\sqrt{\frac{-g_{tt}}{g_{\phi\phi}}}
= \frac{1}{\sin \theta} 
\left( 
\frac{f(r)}{h(r)} 
\right)^{1/2}
\end{equation}
Determining the photon sphere radius corresponds to locating stationary points of $H$ via:
\begin{equation}\label{ph8}
\frac{dH}{dr} = 0.
\end{equation}
To establish a topological framework, consider a vector field $\boldsymbol{\varphi} = (\varphi^{r_h}, \varphi^\theta)$, whose components are:
\begin{equation}\label{ph9}
\varphi^{r_h} = \sqrt{g(r)} \frac{dH}{dr}, \quad \varphi^\Theta = \frac{\partial_\theta H}{\sqrt{h(r)}}.
\end{equation}
This vector can be expressed in polar form as:
\begin{equation}\label{ph10}
\varphi = |\varphi| e^{i \Theta}, \quad \text{where} \quad |\varphi| = \sqrt{(\varphi^{r_h})^2 + (\varphi^\Theta)^2},
\end{equation}
or equivalently in the complex plane by:
\begin{equation}\label{ph11}
\varphi = \varphi^{r_h} + i \varphi^\Theta.
\end{equation}
By normalizing $\boldsymbol{\varphi}$, one obtains the unit vector:
\begin{equation}\label{ph12}
n^a = \frac{\varphi^a}{|\varphi|}, \quad \text{with} \quad \varphi^1 = \varphi^{r_h}, \quad \varphi^2 = \varphi^\Theta.
\end{equation}
To deepen the topological classification, scalar fields $\phi^{r_h}$ and $\phi^\Theta$ are introduced, which further characterize the nature of photon spheres from a topological perspective.
\begin{equation}\label{ph13}
\phi^{r_h}=\frac{\csc (\theta ) \left(-\left(r_h^2 \left(m^2 \left(4 c_2 c^2+\text{cc}_1 r_h\right)+4\right)\right)-2 M e^{-\frac{k}{2 r_h}} (k-6 r_h)\right)}{4 r_h^4},
\end{equation}
and
\begin{equation}\label{ph14}
\phi^\Theta=-\frac{\cot (\theta ) \csc (\theta ) \sqrt{m^2 \left(c_2 c^2+\frac{\text{cc}_1 r_h}{2}\right)-\frac{2 M e^{-\frac{k}{2 r_h}}}{r_h}+\frac{r^2}{l^2}+1}}{r_h^2}.
\end{equation}
According to Fig.~\ref{fig1}, the influence of the massive gravity parameters significantly affects the roots of the metric function. Specifically, the nature of these roots characterizes the black hole geometry as follows: if the metric function has no roots, the solution corresponds to a naked singularity (bare singularity); if there is a single root, the black hole is extremal; and if there are two roots, the spacetime contains both an inner horizon and an outer horizon, with the latter identified as the event horizon. In this work, we investigate the properties of nonlinear charged Anti-de Sitter (AdS) black holes within the framework of massive gravity, focusing on the regime where the charge-to-mass ratio satisfies 
$
\frac{q}{m} > \left(\frac{Q}{M}\right)_{\text{ext}},
$
where \(\left(\frac{Q}{M}\right)_{\text{ext}}\) denotes the charge-to-mass ratio of a large extremal black hole. To this end, we present a series of figures that test the validity of the Weak Gravity Conjecture (WGC) and the Weak Cosmic Censorship Conjecture (WCCC), providing evidence supporting the WGC through the existence of photon spheres (PSs) and the observation of the Aschenbach phenomenon. Generally, a charged black hole can have either two event horizons if 
$
\frac{Q^2}{M^2} \leq 1,
$
or no event horizon if 
$
\frac{Q^2}{M^2} > 1,
$
the latter scenario resulting in a naked singularity that contradicts the WCCC. However, when considering nonlinear charged AdS black holes in massive gravity, we observe that event horizons can still exist in certain parameter regions even when 
$
\frac{Q^2}{M^2} > 1,
$
thereby ensuring that the singularity remains hidden behind a horizon. This result indicates that both the WGC and WCCC are simultaneously satisfied in these cases. Moreover, this framework allows for the exploration of other related phenomena. Notably, in the extremal limit of the black hole, these conjectures remain valid, with the black hole preserving its event horizon. Using the data presented in Tables~\ref{P1} and~\ref{P2}, we identify parameter regimes for the nonlinear electrodynamics and massive gravity sectors in which the WCCC and WGC are compatible. Furthermore, we summarize the findings illustrated in Figs.~\ref{fig2} through \ref{fig5} in Tables~\ref{P3} and~\ref{P4}. From these, it is evident that when the parameter \(c_1\) is positive, the system exhibits conditions under which a spherical photon orbit outside an extremal black hole becomes unstable. This instability is consistent with the predictions of the WGC, providing strong support for the conjecture and its potential observational implications. Conversely, if the parameter \(c_1\) is negative, two photon spheres arise outside the black hole: one stable and one unstable. This scenario suggests a connection between the Aschenbach phenomenon and the WGC, which warrants further detailed investigation. Overall, these results contribute to a deeper understanding of how nonlinear electrodynamics and massive gravity parameters influence black hole horizons and the interplay between fundamental conjectures in gravitational physics.

\begin{table}[h!]
\centering
\setlength{\arrayrulewidth}{0.4mm} 
\setlength{\tabcolsep}{2.8pt} 
\arrayrulecolor[HTML]{000000} 
\begin{tabular}{|>{\centering\arraybackslash}m{2cm}|>{\centering\arraybackslash}m{2cm}|>{\centering\arraybackslash}m{2cm}|>{\centering\arraybackslash}m{2cm}|>{\centering\arraybackslash}m{2cm}|>{\centering\arraybackslash}m{2cm}|>{\centering\arraybackslash}m{2cm}|>{\centering\arraybackslash}m{2cm}|}
\hline
\rowcolor[HTML]{9FC5E8} 
\textbf{$c_1$} & \textbf{$c_2$} & \textbf{$Q_e$} & \textbf{$r_{e}$} & \textbf{$M_{e}$} & \textbf{$(Q/M)_{e}>1$} & \textbf{$WGC-WCCC$}\\ \hline
\rowcolor[HTML]{EAF4FC} 
-1 & -1 & 1 & 0.6962 & 0.6981 & 1.4325 & $\checkmark$ \\ \hline
-1 & 1 & 1 & 0.5509 & 0.9086 & 1.1006 & $\checkmark$ \\ \hline
\rowcolor[HTML]{EAF4FC}
1 & -1 & 1 & 0.5603 & 0.7668 & 1.3041 & $\checkmark$ \\ \hline
1 & 1 & 1 & 0.4832 & 0.9546 & 1.0476 & $\checkmark$ \\ \hline

\end{tabular}
\caption{The condition for consistency WGC-WCCC for $\ell=3$, $m=0.5$ and $c=1$.}
\label{P2}
\end{table}

\begin{table}[h!]
\centering
\setlength{\arrayrulewidth}{0.4mm} 
\setlength{\tabcolsep}{2.8pt} 
\arrayrulecolor[HTML]{000000} 
\begin{tabular}{|>{\centering\arraybackslash}m{2cm}|>{\centering\arraybackslash}m{2cm}|>{\centering\arraybackslash}m{2cm}|>{\centering\arraybackslash}m{2cm}|>{\centering\arraybackslash}m{2cm}|>{\centering\arraybackslash}m{2cm}|>{\centering\arraybackslash}m{2cm}|>{\centering\arraybackslash}m{2cm}|}
\hline
\rowcolor[HTML]{9FC5E8} 
\textbf{$c_1$} & \textbf{$c_2$} & \textbf{$PS$} & \textbf{$q/m>(Q/M)_{e}$} & \textbf{$PS-WGC$}\\ \hline
-1 & -1 & 0 & 1.1985 & $\times$ \\ \hline
\rowcolor[HTML]{EAF4FC} 
-1 & 1 & 0 & 1.1873 & $\times$ \\ \hline
1 & -1 & -1 & 1.1953 & $\checkmark$ \\ \hline
\rowcolor[HTML]{EAF4FC} 
1 & 1 & -1 & 1.1842 & $\checkmark$ \\ \hline
\end{tabular}
\caption{The condition for consistency PS-WGC for $\ell=3$, $m=0.1$ and $c=1$.}
\label{P3}
\end{table}
\begin{table}[h!]
\centering
\setlength{\arrayrulewidth}{0.4mm} 
\setlength{\tabcolsep}{2.8pt} 
\arrayrulecolor[HTML]{000000} 
\begin{tabular}{|>{\centering\arraybackslash}m{2cm}|>{\centering\arraybackslash}m{2cm}|>{\centering\arraybackslash}m{2cm}|>{\centering\arraybackslash}m{2cm}|>{\centering\arraybackslash}m{2cm}|>{\centering\arraybackslash}m{2cm}|>{\centering\arraybackslash}m{2cm}|>{\centering\arraybackslash}m{2cm}|}
\hline
\rowcolor[HTML]{9FC5E8} 
\textbf{$c_1$} & \textbf{$c_2$} & \textbf{$PS$} & \textbf{$q/m>(Q/M)_{e}$} & \textbf{$PS-WGC$}\\ \hline
-1 & -1 & 0 & 1.4325 & $\times$ \\ \hline
\rowcolor[HTML]{EAF4FC} 
-1 & 1 & 0 &  1.1006 & $\times$ \\ \hline
1 & -1 & -1 &1.3041 & $\checkmark$ \\ \hline
\rowcolor[HTML]{EAF4FC} 
1 & 1 & -1 &  1.0476 & $\checkmark$ \\ \hline
\end{tabular}
\caption{The condition for consistency PS-WGC for $\ell=3$, $m=0.5$ and $c=1$.}
\label{P4}
\end{table}
\section{Aschenbach-Like Effect and Extremality}     
As inferred from the analysis conducted so far, the primary goal of this study is to demonstrate the existence of a class of charged black holes that, throughout their evolution, naturally tend toward either a super-extremal charge state or the extremal configuration. This progression arises due to a defined mechanism involving mass-energy absorption and radiation, resulting in an increasing charge-to-mass ratio that ultimately surpasses unity—a scenario conducive to the realization of the Weak Gravity Conjecture (WGC). Our observations, in line with those reported in WGC-related studies, indicate that such gravitational structures—Despite extremality and exhibiting near-zero Hawking temperature—retain fundamental features such as the event horizon and photon sphere even at the threshold of extremality.
On one hand, extremality, owing to its unique features (e.g. vanishing Hawking temperature), represents a physically unstable regime that black holes typically move away from. On the other hand, the tendency to preserve pre-extremal configurations despite their inherent instability could indicate an intrinsic tendency, meaning that rather than evolving into an unfamiliar or undefined gravitational regime (swampland), the black hole may shed excess charge through momentary electromagnetic dominance over gravity, thereby reestablishing its equilibrium and continuing its standard evolution as predicted by general relativity \cite{401}. This process encapsulates the essence of WGC in a natural framework.
Recent investigations suggest that massive black holes, including the one analyzed here, could serve as potential candidates for exhibiting the so-called Aschenbach-like phenomenon \cite{403}. 
Originally identified in Kerr black holes \cite{404}, this phenomenon revealed a non-monotonic behavior in the angular velocity profile of rotating particles. Classical expectations predict a smooth decrease in orbital speed with increasing radial distance; however, observations from Kerr geometry revealed deviations driven by frame-dragging and intense gravitational curvature effects.
Subsequent studies extended the possibility of observing similar behavior in non-rotating black holes \cite{405,406}, attributing the phenomenon to the emergence of a stable photon sphere outside the event horizon \cite{46m}. These findings confirm that spacetime curvature alone, if the gravitational architecture allows for the presence of potential minima beyond the horizon, can generate such dynamical anomalies. Given our earlier discussion showing black holes' resilience in maintaining their pre-extremal conditions, an intriguing question arises:
If Aschenbach-like properties are a behavioral characteristic of our model under normal conditions, do these properties also persist in the extreme regime?
To address this question, we explore the presence and characteristics of the Aschenbach-like phenomenon under extremal conditions in our specific model.
Since, according to studies \cite{403,46m}, the possibility of demonstrating this phenomenon in non-rotating black holes depends on the emergence of a potential minimum (a stable photon sphere) outside the event horizon, we first turn to the photon sphere model, considering equations ~\ref{ph7} and ~\ref{ph9}.
\begin{figure}[]
 \begin{center}
 \subfigure[]{
 \includegraphics[height=4cm,width=5cm]{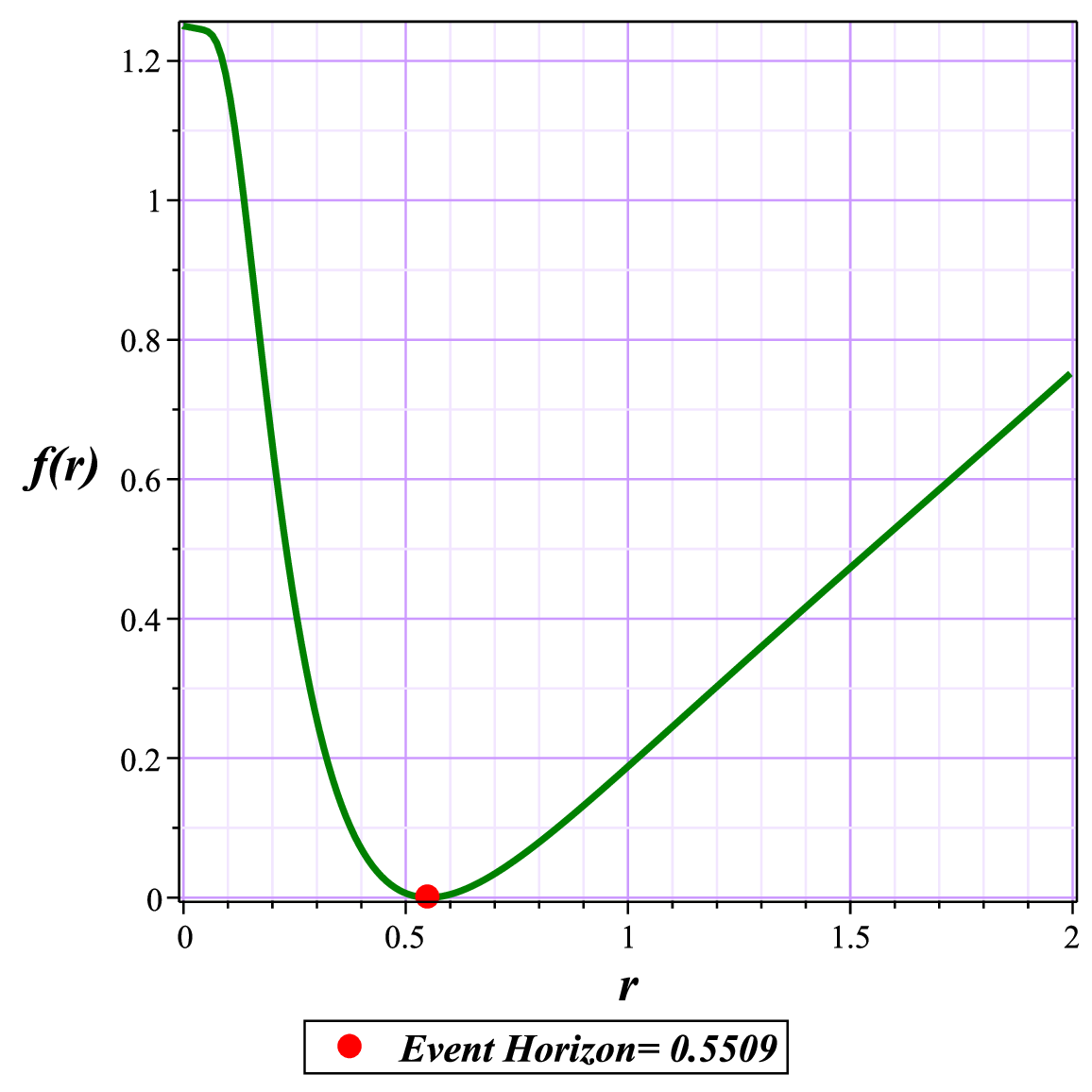}
 \label{6a}}
 \subfigure[]{
 \includegraphics[height=4cm,width=5cm]{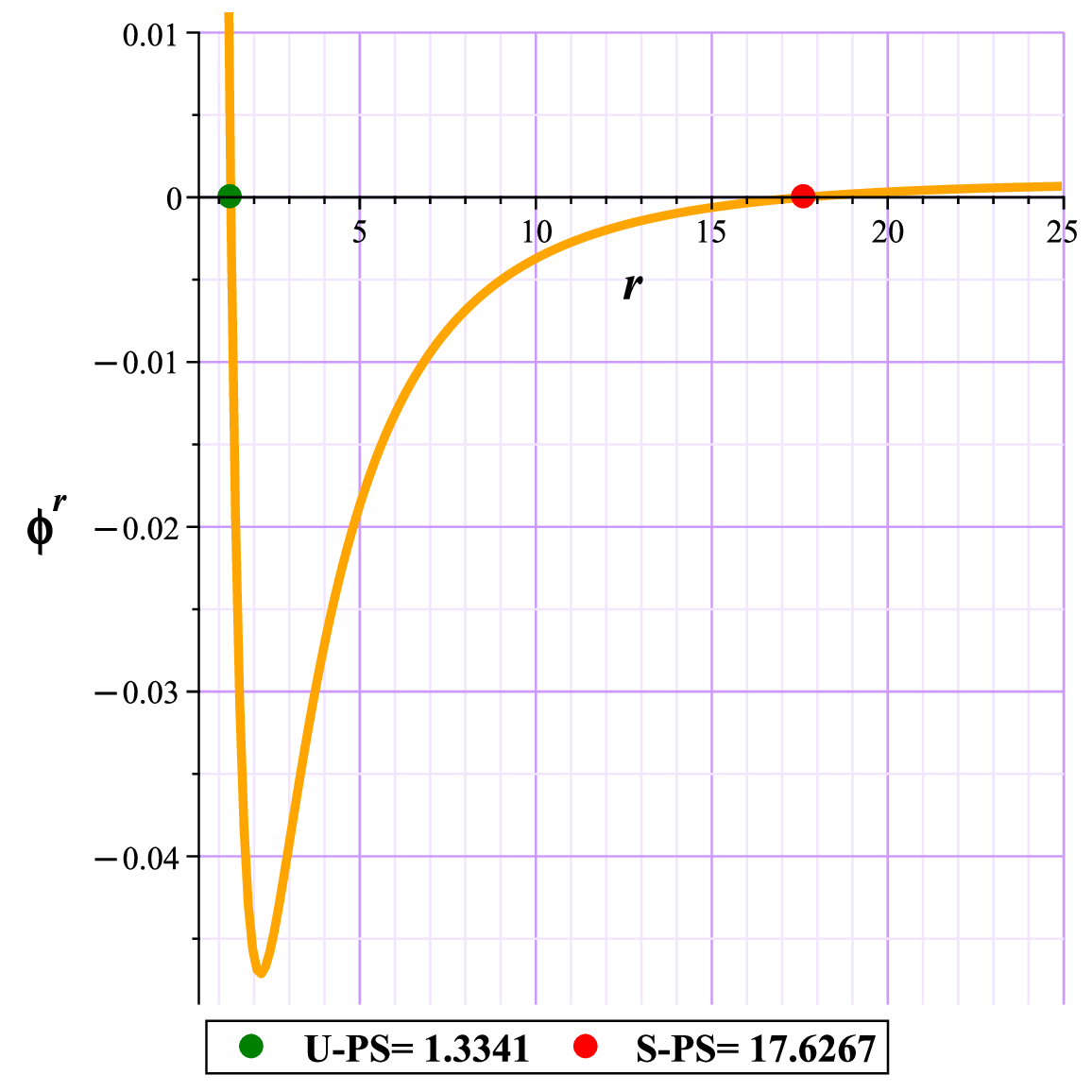}
 \label{6b}}
\subfigure[]{
 \includegraphics[height=4cm,width=5cm]{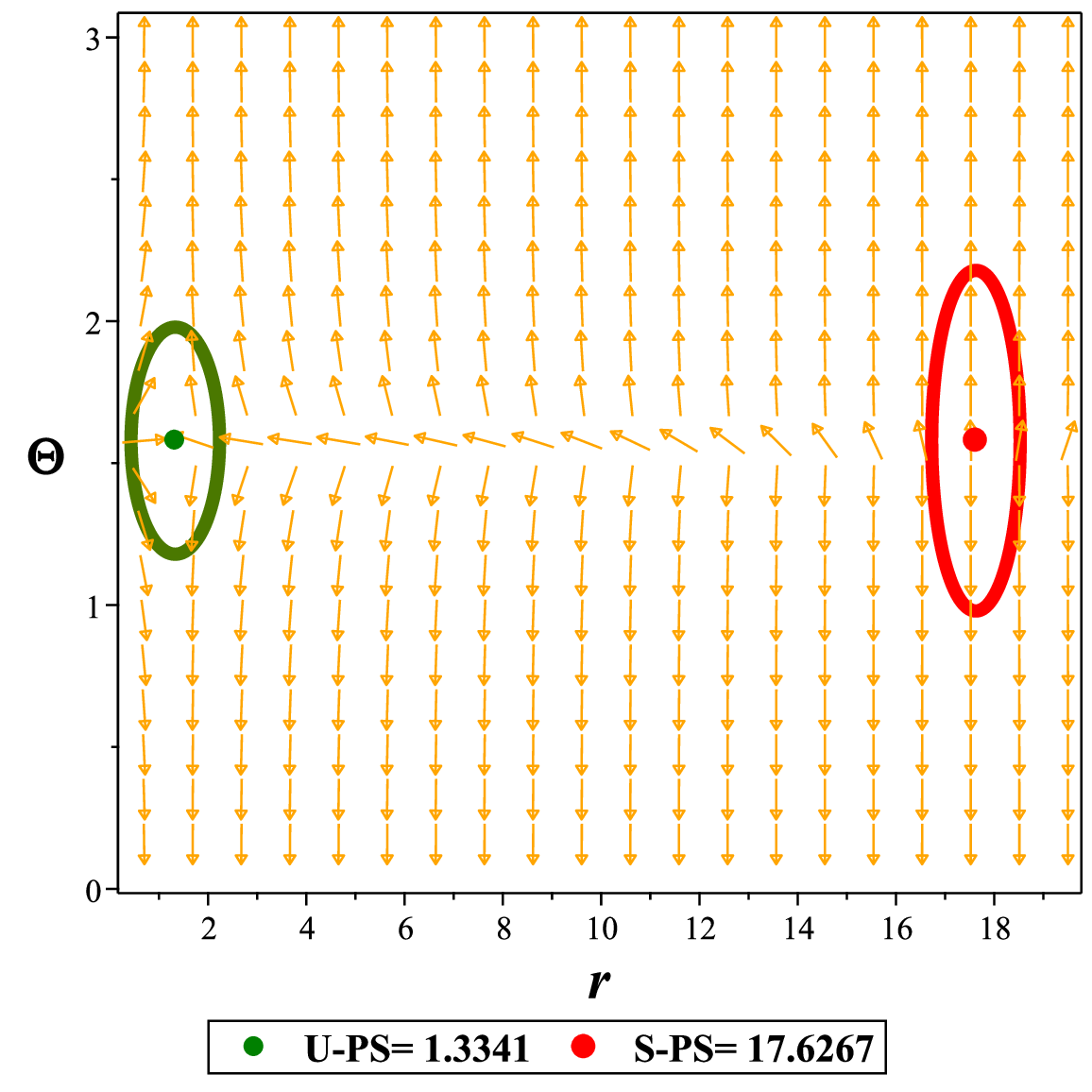}
 \label{6c}}
 \caption{\small{(6a): Metric function With $ M=0.90861,l = 3, k = 1.10058, c = 1, c_{1} = -1, c_{2} = 1, m_{g} = 0.5$, (6b): The photon spheres location at $ (r, \theta)=(1.3341, 1.57)$ and $ (r, \theta)=(17.6267, 1.57)$ (6c): The unstable photon sphere(U-PS) with topological charge $-1$ (green) and stable photon sphere(S-PS) with topological charge $+1$ (red) for Non-linear charged AdS black hole in massive gravity }}
 \label{1}
\end{center}
\end{figure}
As shown in Figure (\ref{1}), while this model actually exhibits black hole behavior (\ref{6a}), two distinct photon spheres—an unstable photon sphere and a stable photon sphere—both appear outside the event horizon (\ref{6b}) ,(\ref{6c}).
Now, to analyze the non-monotonicity of the angular velocity($\Omega$) behavior, we first need some practical relations, which we will have according to Eqs. (~\ref{m9} and ~\ref{m11}) \cite{403}:
\begin{equation}
\begin{split}
&\beta =-r^{2} \Omega^{2}+f(r),\\
&E=\frac{f(r)}{\sqrt{\beta}},\\
&L=-\frac{r^{2} \Omega}{\sqrt{\beta}},\\
&\Omega=\sqrt{\frac{f \! \left(r \right)-\beta}{r^{2}}}.\\
\label{18As}
\end{split}
\end{equation}
where $E$ , $L$ and $f(r)$ are the energy and angular momentum  and metric function respectivly. An important point to note regarding ~\ref{18As} is the value of the function $\beta$. In regions where $\beta$ is negative, the energy and angular momentum become imaginary, effectively prohibiting these regions from the presence of Time-like Circular Orbits(TCOs).
\begin{figure}[]
 \begin{center}
 \subfigure[]{
 \includegraphics[height=4cm,width=5cm]{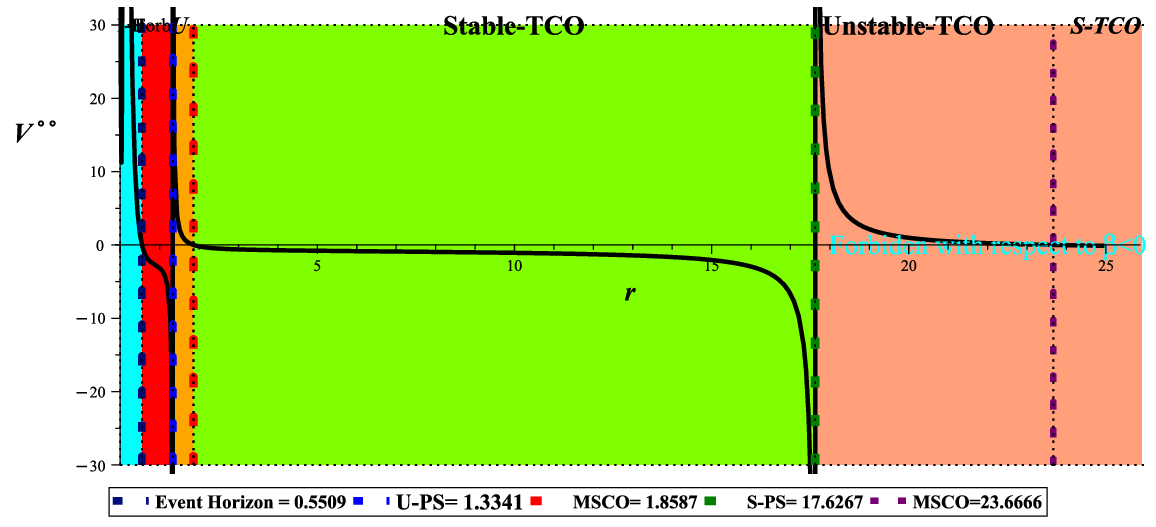}
 \label{7a}}
 \subfigure[]{
 \includegraphics[height=4cm,width=5cm]{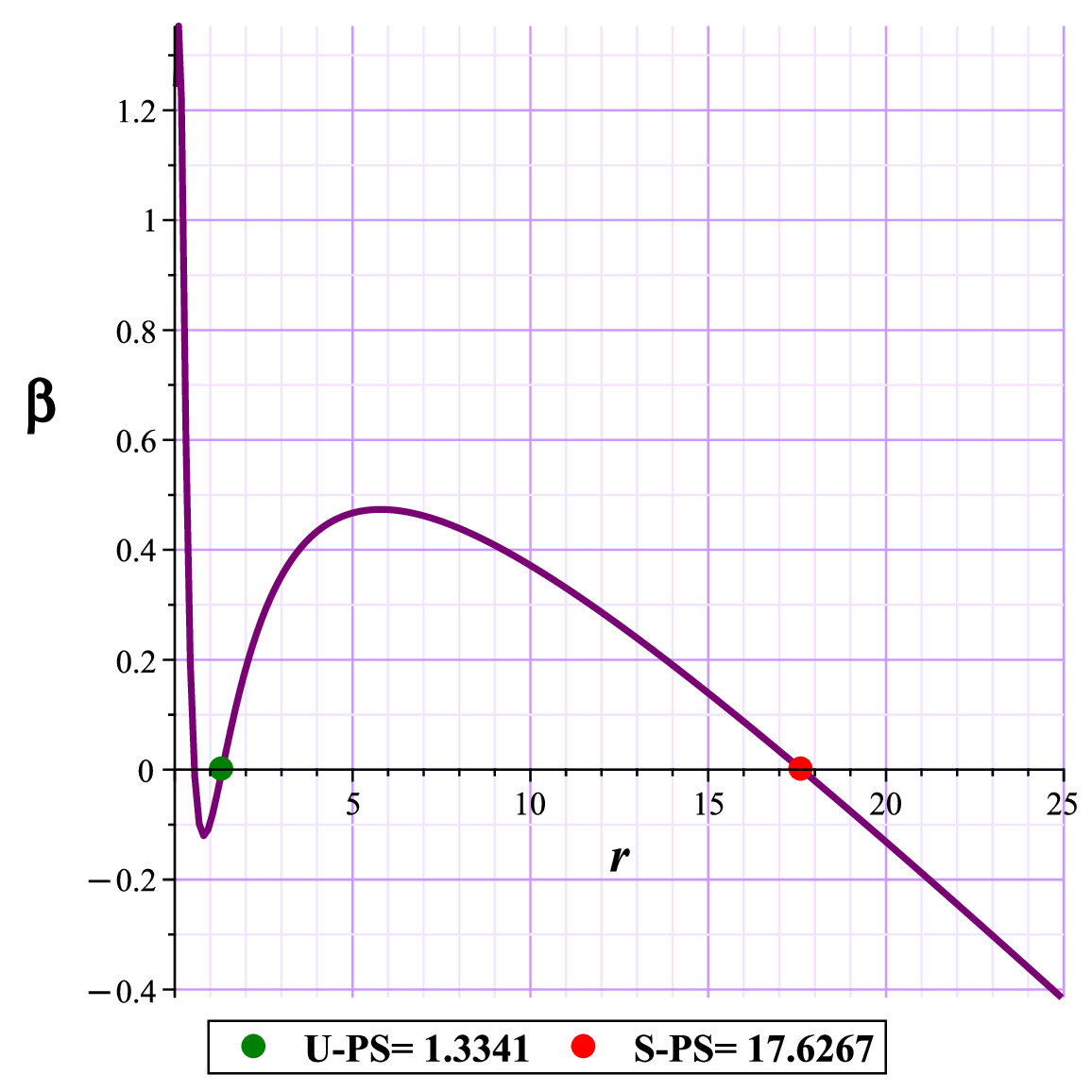}
 \label{7b}}
 \subfigure[]{
 \includegraphics[height=4cm,width=5cm]{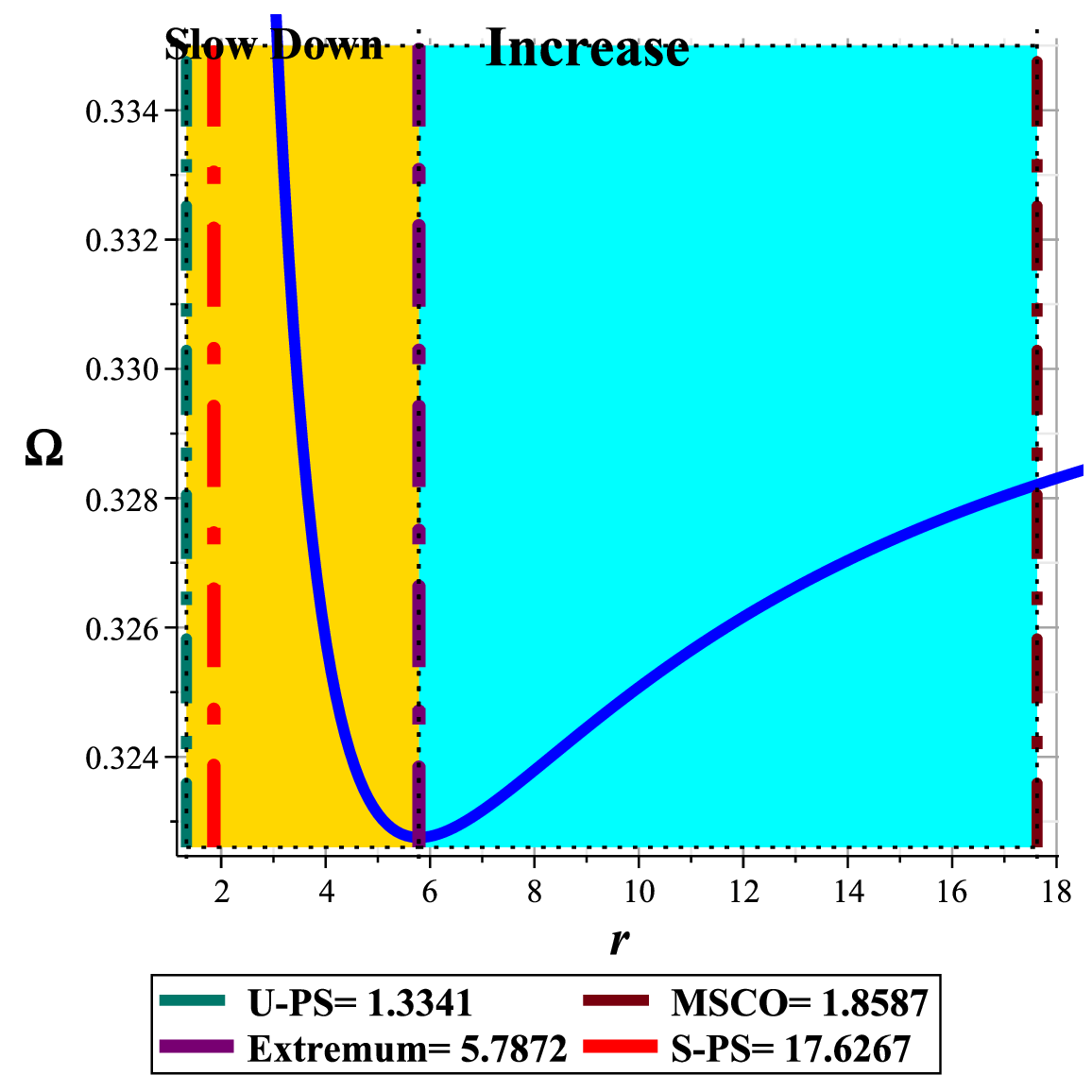}
 \label{7c}}
   \caption{\small{Fig (7a): MSCO localization and space classification    (7b):$\beta$ diagram (7c): Angular velocity VS r with $ M=0.90861,l = 3, k = 1.10058, c = 1, c_{1} = -1, c_{2} = 1, m_{g} = 0.5$ for Non-linear charged AdS black hole in massive gravity }}
 \label{2}
\end{center}
\end{figure}
As shown in Fig. (\ref{7b}), the region allowed for the emergence of TCOs is between two photon spheres (\ref{7a}). \\
In this region the angular velocity shows a decreasing trend along the unstable TCOs originating from the edge of the unstable photon sphere. However, this uniform decrease does not continue indefinitely. Approaching a minimum at \(r = 5.7872\) and continuing towards the stable photon sphere, the velocity starts to increase, leading to a globally non-uniform profile. This slope inversion is exactly the kind of behavior we intended to detect as a sign of Aschenbach-like dynamics. \\
This appears to constitute further confirmation of the earlier proposition. It suggests that the underlying structure exhibits a persistent tendency to conserve its prior configurations under all circumstances, actively maintaining its dominant pre-extremal characteristics even as the system approaches regimes of instability, such as those governing the extremal configuration.
\section{Conclusion} \label{sec5}
In this work, we have explored the nonlinear charged AdS black hole solutions within the framework of massive gravity, with particular emphasis on the persistence of the Aschenbach effect—a unique relativistic phenomenon strongly linked to the presence and properties of photon spheres. Motivated by prior research supporting the WGC through approaches such as WCCC and photon sphere analysis, we extended the investigation of the Aschenbach effect beyond its traditional domain of sub-extremal non rotating black holes. Our analysis demonstrates that the Aschenbach effect remains valid not only up to the extremal limit of black holes but also continues to manifest consistently as the system crosses into the superextremal regime. This finding is significant as it suggests that the characteristic geometrical and dynamical features governing the effect—rooted in the structure of spacetimes and photon spheres—are remarkably robust, surviving even as the classical boundaries of black hole solutions are approached and exceeded. The persistence of the Aschenbach effect beyond the extremal threshold challenges the conventional understanding of black hole physics and opens a novel perspective on the behavior of ultra-compact objects in extreme gravitational fields. It implies that some fundamental relativistic signatures traditionally associated with stable black hole configurations may extend into regimes that blur the line between black holes and naked singularities or other exotic compact objects.

This work lays the groundwork for further exploration into the interplay between massive gravity theories, nonlinear electromagnetic fields, and relativistic phenomena near black holes. In particular, it invites future studies to investigate the implications for the Weak Gravity Conjecture and Weak Cosmic Censorship in these borderline regimes, potentially offering new insights into the ultimate limits of general relativity and the nature of gravitational collapse. Overall, our results enrich the understanding of how relativistic effects emerge and evolve in highly curved spacetimes, highlighting the resilience of key physical processes even under extreme conditions. This contributes not only to theoretical advancements in black hole physics but also to potential observational prospects, where signatures like the Aschenbach effect might provide novel windows into the physics of rotating black holes in both astrophysical and fundamental contexts.
\section{Appendix}
The work of Saeed Noori Gashti is supported by the Iran National Science Foundation (INSF). This work is based upon research funded
by Iran National Science Foundation (INSF) under project No.4038260

\end{document}